\documentclass{article}
\usepackage{spconf,amsmath,graphicx}

\usepackage{cite}
\usepackage{acronym}
\usepackage{amssymb,amsfonts}
\usepackage{algorithmic}
\usepackage{graphicx}
\usepackage{xcolor}
\usepackage{url}
\usepackage{balance}
\usepackage{subfig}
\usepackage{tabularx}
\usepackage{booktabs,multirow} 
\usepackage{amssymb}
\usepackage{placeins}
\usepackage{pifont}
\usepackage{float}
\usepackage{multirow}
\usepackage{caption}

\usepackage[T1]{fontenc}
\usepackage[colorlinks=true,citecolor=blue,allcolors=blue]{hyperref}

\acrodef{HRTF}{Head-Related Transfer Function}
\acrodef{HRIR}{Head-Related Impulse Response}
\acrodef{TrF}{Transfer Function}
\acrodef{TSE}{Target Speaker Extraction}
\acrodef{TF}{Time-Frequency}
\acrodef{GMM}{Gaussian Mixture Model}
\acrodef{CGMM}{Complex GMM}
\acrodef{WDO}{W-Disjoint Orthogonality}
\acrodef{MLE}{Maximum Likelihood Estimation}
\acrodef{MSE}{Mean Square Error}
\acrodef{RMSE}{Root Mean Square Error}
\acrodef{EM}{Expectation-Maximization}
\acrodef{REM}{Recursive EM}
\acrodef{CREM}{Capp{\'e} and Moulines REM}
\acrodef{RIR}{Room Impulse Responses}
\acrodef{STFT}{Short-Time Fourier transform}
\acrodef{ReLU}{Rectified Linear Unit}
\acrodef{DOA}{Direction of Arrival}
\acrodef{TDOA}{Time Difference of Arrival}
\acrodef{SI-SDR}{Scale Invariant Signal to Distortion Ratio }
\acrodef{FFT}{Fast Fourier Transform}
\acrodef{BSS}{Blind Source Separation}
\acrodef{SSL}{Sound Source Localization}
\acrodef{GCC}{Generalized Cross Correlation }
\acrodef{MUSIC}{Multiple Signals Classification}
\acrodef{SRP-PHAT}{Steered Response Power with Phase Transform}
\acrodef{MAE}{Mean Absolute Error}
\acrodef{WSJ}{Wall Street Journal}
\acrodef{SOFA}{Spatially Oriented Format for Acoustics}
\acrodef{SR}{Sampling Rate}
\acrodef{SNR}{Signal to Noise Ratio}
\acrodef{IR}{Impulse Response}
\acrodef{PESQ}{Perceptual Evaluation of Speech Quality}
\acrodef{MOS}{Mean Opinion Score}
\acrodef{ILD}{Interaural Level Difference}
\acrodef{ITD}{Interaural Time Difference}
\acrodef{STOI}{Short-time Objective Intelligibility}
\acrodef{ATF}{Acoustic Transfer Function}
\acrodef{RTF}{Relative Transfer Function}
\acrodef{RT}{Reverberant Time}
\acrodef{PRP}{Pair-Wise Relative Phase Ratio}
\acrodef{DNN}{Deep Neural Network}
\acrodef{DL}{Deep Learning}
\acrodef{dB}{Decibel}
\acrodef{NN}{Neural Network}
\acrodef{FC}{Fully Connected}
\acrodef{SIR}{Signal-to-Interference Ratio}
\acrodef{p.d.f.}{Probability Density Function}
\acrodef{LCMV}{Linearly Constrained Minimum Variance}
\acrodef{BLCMV}{Binaural Linearly Constrained Minimum Variance}
\acrodef{LS}{LibriSpeech}
\acrodef{MLS}{Multilingual LibriSpeech}
\acrodef{MIMO}{Multi input Multi output}
\acrodef{Bi-TSE}{Binaural Target Speaker Extraction}
\acrodef{Bi-TSE-HRTF}{Binaural Target Speaker Extraction using HRTFs}
\acrodef{RI}{Real-Imaginary}
\acrodef{MOS}{Mean Opinion Score}

\DeclareMathOperator*{\argmax}{arg\,max}
\DeclareMathOperator*{\argmin}{arg\,min}

\title{Unfolded Expectation-Maximization Neural Network for Speaker Localization}
\name{Rina Veler and Sharon Gannot}
\address{Faculty of Engineering, Bar-Ilan University, 
Ramat-Gan, Israel \\
\{velerri,sharon.gannot\}@biu.ac.il}
\begin{document}
\ninept
\maketitle

\begin{abstract}
We propose an interpretable Unfolded \ac{EM}  Network for robust speaker localization. By embedding the iterative EM procedure within an encoder–EM–decoder architecture, the method mitigates initialization sensitivity and improves convergence. Experiments show superior accuracy and robustness over the classical Batch-\ac{EM} in reverberant conditions.

\end{abstract}

\vspace{-3pt}
\section{Introduction}
\vspace{-3pt}
\ac{SSL} is a fundamental task in modern audio applications, from autonomous robots to consumer electronics. Estimating the \ac{DOA} in realistic environments is challenging due to reverberation and noise, and traditional methods, e.g., \ac{SRP-PHAT}, \ac{MUSIC}, and \ac{MLE}, often struggle, especially with multiple sources. This has motivated a shift toward \ac{DNN}-based approaches, now widely used for robust \ac{SSL} in adverse acoustic conditions \cite{Grumiaux2022Survey,Chakrabarty2019MultiSpeaker}. However, purely data-driven \acp{DNN} lack interpretability, whereas classical iterative algorithms remain more transparent. Algorithm unrolling (or unfolding) bridges this gap \cite{Gregor2010Learning,Monga2021Algorithm} by casting algorithmic iterations as differentiable network layers, yielding parameter-efficient, interpretable models that require less training data.
We adopt unfolded networks to estimate static multi-speaker positions using an \ac{MLE} formulation based on a \ac{CGMM}. The observations are \acp{PRP}, with each Gaussian mean representing a location-dependent \ac{PRP} vector. We integrate the classical \ac{EM} procedure—previously used for \ac{PRP} clustering \cite{Schwartz2014}—into an unfolded architecture, yielding a robust and interpretable estimation framework. This follows prior success with \ac{EM}-based unrolling in image clustering and segmentation \cite{Greff2017NeuralEM,Pu2023Deep}.


\vspace{-6pt}
\section{Problem Formulation}
\vspace{-3pt}
\label{Problem Formulation }

Consider an array of $M$ microphone pairs acquiring $S$ speakers with overlapping activities in a reverberant enclosure. The analysis is carried out in the \ac{STFT} domain. We operate under the \ac{WDO} assumption, where each \ac{TF} bin is solely dominated by a single source \cite{Rickard2002WDO}. The number of speakers $S \ge 1$, is assumed a priori known. Therefore, the signal measured at the $j$th microphone of the pair $m$, where $j = 1,2$ and $m = 1,\ldots,M$, is modeled by:
\begin{equation}
z_{m,j}(t,k) = a_{sm}^j(t,k) \cdot v_s(t,k) + n_m^j(t,k)
\end{equation} 
where $t = 0,\dots,T-1$ is the time-frame index and $k=0,\dots,K-1$ denotes the frequency bin. 
The \acp{ATF} describing the propagation from the position of speaker $s$ to microphone $j$ of pair $m$ 
are denoted by $a_{sm}^j(t,k)$, and $n_m^j(t,k)$ denotes additive noise.
In low-reverberation environments, the \ac{ATF} can be approximated by the direct-path 
component, where the speaker and microphone positions, $\mathbf{p}_s$ and $\mathbf{p}_m^j$, 
respectively, determine both the amplitude attenuation and the phase shift.
Rather than using the raw measurements ${z}_{m,j}(t,k)$ directly, we employ the \ac{PRP} as the localization feature.
The \ac{PRP} vector is formed by concatenating the normalized complex-valued ratios from all $M$ microphone pairs:
\begin{equation}
\boldsymbol{\phi}(t,k) = \left[\phi_1(t,k), \dots, \phi_M(t,k)\right]^\top,
\end{equation}
where the individual \ac{PRP} for pair $m$ is defined as
\begin{equation}
\phi_m(t,k) \triangleq \frac{z_{m,2}(t,k)}{z_{m,1}(t,k)} \cdot \frac{|z_{m,1}(t,k)|}{|z_{m,2}(t,k)|}.
\end{equation}
The resulting vector $\boldsymbol{\phi}(t,k)$ constitutes the \textit{observed data} in the subsequent \ac{EM}-based formulation.
We adopt the \ac{CGMM} probability model from \cite{Schwartz2014}, with several modifications.
In our formulation, the number of Gaussian components is set to the number of speakers $S$, whereas in \cite{Schwartz2014} it corresponds to the number of candidate source positions in the environment, typically a much larger set. In that formulation, speaker positions are inferred as the means of the Gaussians with the highest mixture weights.
Under the \ac{WDO} assumption in the \ac{STFT} domain, each \ac{TF} bin of $\boldsymbol{\phi}(t,k)$ is associated with a single active source, leading to the following probabilistic model:
\begin{equation}
\boldsymbol{\phi}(t,k) \sim \sum_{s} \psi_s \cdot
\mathcal{N}^c\left(\boldsymbol{\phi}(t,k);
\tilde{\boldsymbol{\phi}}^k(\mathbf{p}_s), \mathbf{\Sigma}_s\right),
\end{equation}
where $\psi_s$ denotes the prior probability of speaker $s$ being active and $\mathbf{\Sigma}_s$ is the covariance matrix.
The mean of each Gaussian, $\tilde{\boldsymbol{\phi}}^k(\mathbf{p}_s)$, represents the expected \ac{PRP} generated by a speaker at position $\mathbf{p}_s$. This expected mean is derived analytically from the \ac{TDOA}. For pair $m$, the predicted phase ratio is:
\begin{equation}
\tilde{\phi}_m^k(\mathbf{p}_s)
\triangleq
\exp\left(
-j\cdot 2\pi \frac{k}{K}
\frac{
|\mathbf{p}_s - \mathbf{p}_m^2| - |\mathbf{p}_s - \mathbf{p}_m^1|
}{c \cdot T_s}
\right).
\label{eq:PRP}
\end{equation}
Following \cite{Schwartz2014}, we assume independence of the \ac{PRP} features across microphone pairs.
We further simplify each covariance matrix to reflect spatially-white noise, i.e., $\mathbf{\Sigma}_s = \sigma_s^2 \mathbf{I}_M$.
%
%
Finally, the \ac{p.d.f.} of the entire observation set can be written as
\begin{equation}
f(\boldsymbol{\phi}) 
= 
\prod_{t,k} 
\left[
    \sum_{s} 
    \psi_s 
    \prod_{m} 
    \mathcal{N}^c\!\left(
        \phi_m(t,k);\,
        \tilde{\phi}_m^k(\mathbf{p}_s),\,
        \sigma_s^2
    \right)
\right],
\end{equation}
where we assume independence of the \ac{PRP} measurements across all time frames and frequency bins.  

Let $\boldsymbol{\theta} = \left[\boldsymbol{\psi}^\top, (\boldsymbol{\sigma}^2)^\top, \tilde{\boldsymbol{\phi}}^\top\right]$ denote the set of unknown parameters.  
The goal is to solve the \ac{MLE} problem, typically addressed via the \ac{EM} algorithm:
\begin{equation}
\{\hat{\boldsymbol{\psi}},\, \hat{\boldsymbol{\sigma}}^2,\, \hat{\tilde{\boldsymbol{\phi}}}\}
=
\argmax_{\tilde{\boldsymbol{\phi}},\, \boldsymbol{\psi},\, \boldsymbol{\sigma}^2}
\log f\big(\boldsymbol{\phi};\, \tilde{\boldsymbol{\phi}},\, \boldsymbol{\psi},\, \boldsymbol{\sigma}^2\big).
\end{equation}

\section{Proposed Method}
\vspace{-6pt}
\label{Proposed Method}
In this section, we described the proposed method from the conventional \ac{EM} iterations to the unfolded network.

\vspace{3pt}\noindent\textbf{The EM Algorithm:}
\label{The EM Algorithm}
The \ac{EM} algorithm requires specifying the observed data, the hidden variables, and the 
parameters to be estimated. We define the hidden variable $x(t,k,s)$ as an indicator assigning 
each \ac{TF} bin to a single source at a given position.
Given the hidden data, the likelihood of the observations is
\sloppy
\begin{equation}
    f(\mathbf{x}, \boldsymbol{\phi}; \boldsymbol{\theta})  = \prod_{t,k}  \sum_{s} \psi_s x(t,k,s) \prod_{m} \mathcal{N}^c \left(\phi_m(t,k); \tilde{\phi}_m^k(\mathbf{p_s}), \sigma_s^2 \right).
\end{equation}
For implementing the E-step it is sufficient to evaluate $\mu^{(\ell-1)}(t,k,s) \triangleq E\{x(t,k,s) | \boldsymbol{\phi}(t,k); \boldsymbol{\theta}^{(\ell-1)}\}$ given by:
\begin{multline}
\frac{\psi_s^{(\ell-1)} \prod_{m} \mathcal{N}^c \left(\phi_m(t,k); \tilde{\phi}_m^k(p_s)^{(\ell-1)}, (\sigma_s^2)^{(\ell)} \right)}{\sum_{s} \psi_s^{(\ell-1)} \prod_{m} \mathcal{N}^c \left(\phi_m(t,k); \tilde{\phi}_m^k(p_s)^{(\ell-1)}, (\sigma_s^2)^{(\ell-1)} \right)} \label{eq:Estep}
\end{multline}
Maximizing \eqref{eq:Estep} w.r.t.~the parameters $\boldsymbol{\theta}^{(\ell)}$ constitutes the M-step:
\begin{subequations}
\begin{align}
&\psi_s^{(\ell)} = \frac{\sum_{t,k} \mu^{(\ell-1)}(t,k,s)}{T \cdot K} \\
&{\sigma_s^{2}}^{(\ell)} = \frac{\sum_{t,k,m} \mu^{(\ell-1)}(t,k,s) |\phi_m(t,k) - \tilde{\phi}_m^k(\mathbf{p}_s)^{(\ell-1)}|^2}{M \cdot \sum_{t,k} \mu^{(\ell-1)}(t,k,s)} \\
&\tilde{\phi}_m^k(\mathbf{p}_s)^{(\ell)} = \frac{\sum_{t} \mu^{(\ell-1)}(t,k,s) \cdot \phi_m(t,k)}{\sum_{t} \mu^{(\ell-1)}(t,k,s)}.
\end{align}
\end{subequations}
The final position estimate is obtained by searching for the position $\mathbf{p}$ in the room 
(whose dimensions are assumed known) that minimizes:
\begin{equation}
    \hat{\mathbf{p}}_s 
    = 
    \argmin_{\mathbf{p}}
    \sum_{k,m} 
    \left| 
        \tilde{\phi}_m^k(\mathbf{p}_s)^{(\ell)} 
        - 
        \tilde{\phi}_m^k(\mathbf{p})
    \right|^2,
    \qquad \forall s,
\end{equation}
where $\tilde{\phi}_m^k(\mathbf{p})$ is computed for each candidate location $\mathbf{p}$ 
using~\eqref{eq:PRP}.
To enhance \ac{EM} robustness in multiple-speaker scenarios, we set the total number of \ac{CGMM} clusters 
to $S+1$, i.e., the number of speakers plus one. The additional cluster serves as an 
\textit{outlier} cluster, absorbing \ac{TF} bins that do not correspond to any active speaker, 
thereby preventing corruption of the true speaker position estimates. After convergence, this outlier cluster is removed by identifying the cluster with the highest variance.
\begin{figure}[htbp]
\vspace{-3pt}
    \centering
    \includegraphics[width=0.95\linewidth]{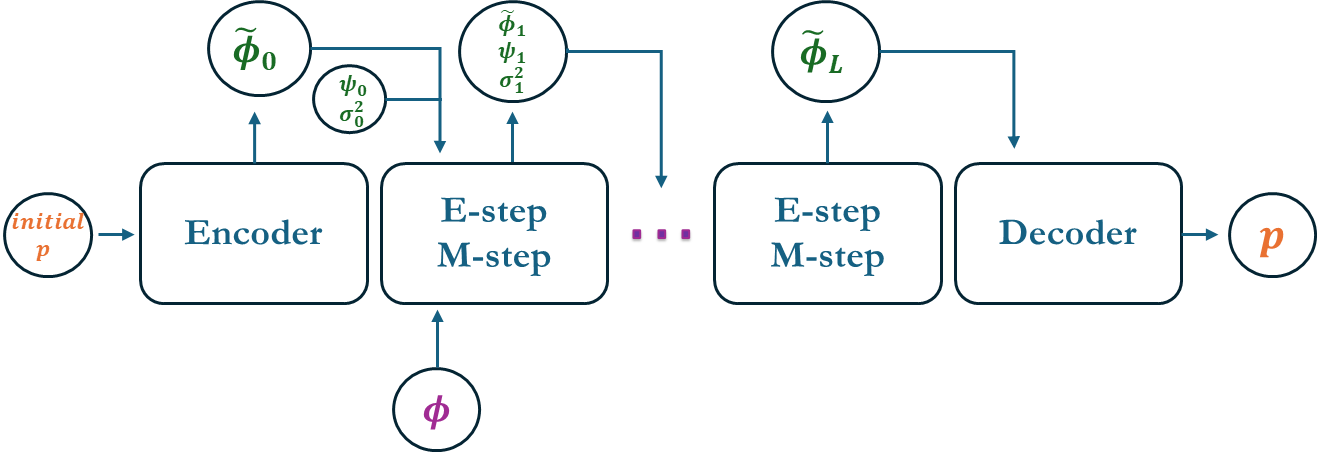}
    \setlength{\belowcaptionskip}{-10pt}
\setlength{\abovecaptionskip}{3pt}
    \caption{Batch EM unfolding neural network architecture.}
    \label{fig:fig1}
\end{figure}

\vspace{2pt}\noindent\textbf{Unfolding Batch EM Neural Network:}
We propose an unfolded \ac{NN} architecture to mitigate the \ac{EM} algorithm’s sensitivity to 
initialization and local optima. The key idea is to embed the iterative \ac{EM} procedure between 
an encoder and a decoder. In the Batch-EM Unfolded Network, a \ac{FC} encoder maps a candidate 
room position to an initial \ac{PRP} vector that initializes the \ac{CGMM} means. The unfolded 
\ac{EM} layers iteratively refine this estimate, and the final \ac{PRP} is passed to a \ac{FC} 
decoder that outputs the speaker positions (\autoref{fig:fig1}). The model is trained using a 
combined loss consisting of a position error term and a \ac{PRP} cosine-distance term, enforcing 
consistency between the refined \ac{PRP} and the theoretical \ac{PRP} at the true positions.
These two losses are weighted as
\begin{equation}
    \mathcal{L}
    = (1-\lambda)\,\text{MSE}(\hat{\mathbf{p}}, \mathbf{p}_{\text{true}})
      + \lambda\,\big(1 - \text{CosSim}(\hat{\boldsymbol{\phi}}, \boldsymbol{\phi}_{\text{true}})\big),
\label{eq:loss function}
\end{equation}
where $\hat{\mathbf{p}}$ and $\mathbf{p}_{\text{true}}$ denote the estimated and true speaker 
positions, and $\hat{\boldsymbol{\phi}}$ and $\boldsymbol{\phi}_{\text{true}}$ the corresponding 
\ac{PRP} vectors.
\vspace{-6pt}
\section{Experimental Study}
\vspace{-6pt}
\label{Experimental study}

\vspace{2pt}\noindent\textbf{Data Generation:}
\label{Signals generation}
We used a synthetic dataset derived from the \ac{WSJ} corpus \cite{Paul1992Design}, simulating two 
static speakers in random rectangular rooms (height: 2.2–2.6 m; width/length: 5–7 m). The recording 
setup included eight microphone pairs (intra-pair distance: 0.2 m). 
The network’s robustness was evaluated under various interference conditions: anechoic and 
reverberant environments with $T_{60}=0.2$~s, temporal overlaps of $25\%, 50\%,$ and $75\%$, 
\ac{SIR} levels of $0$ and $5$~dB, and additive white noise at $\text{SNR}=30$~dB. The dataset 
consisted of $8{,}000$ training and $2{,}000$ validation samples.

\vspace{2pt}\noindent\textbf{Network Architecture and Training:}
\label{Network Architecture:}
We employ a Batch-EM Unfolded Network in which the \ac{EM} procedure is unrolled into $70$ layers. 
Initialization is performed by mapping random room locations to an initial \ac{PRP} vector using a 
\ac{FC} encoder, with $\psi^{(0)}_s=\tfrac{1}{S}$ and ${\sigma_s^2}^{(0)}=1$ for all $s$. The unfolded 
\ac{EM} layers iteratively refine this estimate, and the final \ac{PRP} is mapped to the speakers' 
positions by an \ac{FC} decoder. All \ac{FC} layers use \ac{ReLU} activation.
Because the \ac{PRP} is complex-valued, the \ac{FC} layers operate on concatenated real and 
imaginary parts and then reshape the output back to a complex vector. The model is trained using the 
loss in \eqref{eq:loss function} with $\lambda=0.25$.

\vspace{2pt}\noindent\textbf{Results and Discussion:}
\label{Results}
We evaluated the proposed approach against the Batch-\ac{EM} algorithm with exhaustive spatial 
grid search, using $100$ unseen speech mixtures. Performance was assessed in terms of localization 
accuracy and robustness across different acoustic and interference conditions.
\autoref{tab:results} summarizes the results, averaged over all SIR and overlap settings for each 
environment. Metrics include the \ac{RMSE} between estimated and actual speaker positions and the 
percentage of samples with positional error above $0.5$~m.
\begin{table}[htbp] %
\centering
\setlength{\belowcaptionskip}{-6pt}
\setlength{\abovecaptionskip}{3pt}
\caption{Average Localization Performance: Unfolded Network vs. Batch \ac{EM} Baseline, Averaged over all \ac{SIR} and Overlap Conditions.}
\label{tab:results}
\resizebox{0.98\columnwidth}{!}{
\begin{tabular}{lcccc}
\toprule
\multirow{2}{*}{\textbf{Rever. Level}} & \multicolumn{2}{c}{\textbf{Unfolded \ac{EM} Network}}\rule{0pt}{12pt} & \multicolumn{2}{c}{\textbf{Batch \ac{EM} (Baseline)}} \\
\cline{2-5}
& \textbf{\ac{RMSE} ($\text{m}$)}\rule{0pt}{12pt} & \textbf{Error $> 0.5\text{ m}$ ($\%$)} & \textbf{\ac{RMSE} ($\text{m}$)} & \textbf{Error $> 0.5\text{ m}$ ($\%$)} \\
\midrule
$T_{60} = 0$~s & 0.31\rule{0pt}{12pt} & 15.5 & \textbf{0.25} & \textbf{12.5} \\
$T_{60} = 0.2$~s & \textbf{0.37}\rule{0pt}{12pt} & \textbf{22} & 0.66 & 56 \\
\bottomrule
\end{tabular}}
\vspace{-6pt}
\end{table}
As shown in \autoref{tab:results}, the Batch-EM method performs slightly better in clean 
conditions, since the deterministic \text{Scan-to-Locate} procedure yields an almost exact solution 
when reflections are limited. The unfolded network introduces small approximation errors due to the 
\ac{FC} mappings.
In reverberant settings, however, the proposed Unfolded \ac{EM} Network offers substantially better 
generalization and robustness. It reduces the \ac{RMSE} by about $39\%$ relative to the Batch-\ac{EM} 
baseline and significantly lowers the rate of large localization errors. This improvement results 
from the network’s learned mapping, which compensates for reverberation-induced distortion in the 
\ac{PRP} features.

\vspace{-6pt}
\section{Conclusion}
\vspace{-6pt}
\label{Conclusion}

We developed a novel Unfolded \ac{EM} Network architecture for speaker localization by embedding the iterative \ac{EM} procedure within a learnable network. Our results confirmed that the network achieved significantly superior robustness in reverberant environments, showing an approximate $40\%$ reduction in \ac{RMSE} compared to the Batch-\ac{EM} baseline. 

\balance
\bibliographystyle{IEEEbib}
\bibliography{refs25}

@string{icassp = "Proc. ICASSP"}

@string{icml = "Proc. ICML"}

@string{neurips = "Proc. NeurIPS"}

@article{Schwartz2014,
  author    = {Schwartz, Ofer and Gannot, Sharon},
  title     = {Speaker Tracking Using Recursive EM Algorithms},
  journal   = {IEEE/ACM Transactions on Audio, Speech, and Language Processing},
  volume    = {22},
  number    = {2},
  pages     = {392--402},
  year      = {2014},
  month     = feb,
  doi       = {10.1109/TASLP.2013.2292361},
  publisher = {IEEE}
}

@article{Chakrabarty2019MultiSpeaker,
  author    = {Chakrabarty, Soumitro and Habets, Emanuel A.},
  title     = {Multi-Speaker DOA Estimation Using Deep Convolutional Networks Trained With Noise Signals},
  journal   = {IEEE Journal of Selected Topics in Signal Processing},
  volume    = {13},
  number    = {1},
  pages     = {8--21},
  year      = {2019},
  publisher = {IEEE}
}

@inproceedings{Gregor2010Learning,
  author    = {Gregor, Karol and LeCun, Yann},
  title     = {Learning fast approximations of sparse coding},
  booktitle = {Proceedings of the International Conference on Machine Learning (ICML)},
  pages     = {399--406},
  year      = {2010},
  doi       = {10.5555/3104322.3104374},
  publisher = {ACM}
}

@article{Monga2021Algorithm,
  author    = {Monga, V. and Li, Y. and Eldar, Y. C.},
  title     = {Algorithm Unrolling: Interpretable, Efficient Deep Learning for Signal and Image Processing},
  journal   = {IEEE Signal Processing Magazine},
  volume    = {38},
  number    = {2},
  pages     = {18--44},
  year      = {2021},
  publisher = {IEEE},
  doi       = {10.1109/MSP.2020.3042412} 
}

@inproceedings{Greff2017NeuralEM,
  author    = {Greff, Klaus and van Steenkiste, Sjoerd and Schmidhuber, Juergen},
  title     = {Neural Expectation Maximization},
  booktitle = {Proceedings of the 31st Conference on Neural Information Processing Systems (NeurIPS)},
  pages     = {6694--6704},
  year      = {2017}
}

@article{Pu2023Deep,
  author    = {Pu, Yannan and Sun, Jian and Tang, Niansheng and Xu, Zongben},
  title     = {Deep expectation-maximization network for unsupervised image segmentation and clustering},
  journal   = {Image and Vision Computing},
  volume    = {135},
  pages     = {104717},
  year      = {2023},
}

@inproceedings{Paul1992Design,
  author    = {Paul, Douglas B. and Baker, Janet M.},
  title     = {The design for the Wall Street Journal-based CSR corpus},
  booktitle = {Proceedings of the workshop on Speech and Natural Language},
  pages     = {357--362},
  year      = {1992},
  publisher = {Association for Computational Linguistics},
  organization = {ACL}
}

@article{Grumiaux2022Survey,
  author    = {Grumiaux, Pierre-Amaury and Kiti{\'c}, Sr{\dj}an and Girin, Laurent and Gu{\'e}rin, Alexandre},
  title     = {A survey of sound source localization with deep learning methods},
  journal   = {J. Acoust. Soc. Am.},
  volume    = {152},
  number    = {1},
  pages     = {107--151},
  year      = {2022},
  month     = jul
}

@INPROCEEDINGS{Rickard2002WDO,
  author={Rickard, Scott and Yilmaz, Ozgiir},
  booktitle={IEEE International Conference on Acoustics, Speech, and Signal Processing (ICASSP)}, 
  title={On the approximate {W}-disjoint orthogonality of speech}, 
  year={2002},
  volume={1},
  number={},
  pages={529--532}
}
\end{document}